\documentclass[pre,floats,twocolumn,showpacs,superscriptaddress]{revtex4-1}

\usepackage{graphicx}
\usepackage{rotate}
\usepackage{epsfig}
\usepackage{xcolor}
\usepackage[normalem]{ulem}

\usepackage{amssymb,amsfonts,amsmath}

\newcommand{\etal}{\emph{et al.} }

\definecolor{MyBrick}{rgb}{0.84,0.01,0.01}

\newcommand{\avg}[1]{\langle #1 \rangle}

\begin{document}

\title{Evolutionary dynamics of time-resolved social interactions}

\author{Alessio Cardillo}
\affiliation{Departamento de F\'{\i}sica de la Materia Condensada, Universidad de Zaragoza, E-50009 Zaragoza, Spain}
\affiliation{Institute for Biocomputation and Physics of Complex Systems (BIFI), University of Zaragoza, E-50018 Zaragoza, Spain}

\author{Giovanni Petri}
\affiliation{Institute for Scientific Interchange (ISI), via Alassio 11/c, 10126 Torino, Italy}
 
\author{Vincenzo Nicosia}
\affiliation{School of Mathematical Sciences, Queen Mary University of London, Mile End Road, E14NS London, UK}

\author{Roberta Sinatra}
\affiliation{Center for Complex Network Research and Department of Physics, Northeastern University, Boston, MA 02115, USA}
\affiliation{Center for Cancer Systems Biology, Dana-Farber Cancer Institute, Boston, MA 02115, USA}

\author{Jes\'us G\'omez-Garde\~nes}
\email{gardenes@gmail.com}
\affiliation{Departamento de F\'{\i}sica de la Materia Condensada, Universidad de Zaragoza, E-50009 Zaragoza, Spain}
\affiliation{Institute for Biocomputation and Physics of Complex Systems (BIFI), University of Zaragoza, E-50018 Zaragoza, Spain}

\author{Vito Latora}
\affiliation{School of Mathematical Sciences, Queen Mary University of London, Mile End Road, E14NS London, UK}
\affiliation{Dipartimento di Fisica e Astronomia, Universit\`a di Catania, and INFN, Via S. Sofia 64, I-95123 Catania, Italy}



\begin{abstract}
Cooperation among unrelated individuals is frequently observed in
social groups when their members combine efforts and resources to obtain
a shared benefit that is unachievable by an individual
alone. However, understanding why cooperation arises despite the
natural tendency of individuals towards selfish behavior is still an
open problem and represents one of the most fascinating challenges in
evolutionary dynamics.
Recently, the structural characterization of the networks in
which social interactions take place has shed some light on the
mechanisms by which cooperative behavior emerges and eventually
overcomes the natural temptation to defect. In particular, it has been
found that the heterogeneity in the number of social ties and the
presence of tightly knit communities lead to a significant increase in
cooperation as compared with the unstructured and homogeneous
connection patterns considered in classical evolutionary dynamics.
Here, we investigate the role of social-ties dynamics for the emergence
of cooperation in a family of social dilemmas. Social interactions are
in fact intrinsically dynamic, fluctuating, and intermittent over time,
and they can be represented by time-varying networks. By considering two experimental data sets of human interactions
with detailed time information, we show that the temporal dynamics of
social ties has a dramatic impact on the evolution of cooperation: the
dynamics of pairwise interactions favors selfish behavior.
\end{abstract}


\maketitle


\section{Introduction}

The organizational principles driving the evolution and development of
natural and social large-scale systems, including populations of
bacteria, ant colonies, herds of predators, and human societies, rely
on the cooperation of large populations of unrelated agents
\cite{penni05,penni09,maynard95}.  Even if cooperation seems to be a
ubiquitous property of social systems, its spontaneous emergence is
still a puzzle for scientists, since cooperative behaviors are
constantly threatened by the natural tendency of individuals towards
self-preservation and the never-ceasing competition among agents for
resources and success. The preference for selfishness over cooperation
is also due to the higher short-term benefits that a single (defector)
agent obtains by taking advantage of the joint efforts of cooperating
agents. Obviously, the imitation of such selfish (but rational)
conduct drives the system towards a state in which the higher benefits
associated with cooperation are no longer achievable, with dramatic
consequences for the whole population.  Consequently, the relevant
question to address is why cooperative behavior is so common in nature
and society, and what are the circumstances and the mechanisms that
allow it to emerge and persist.

In recent decades, the study of the elementary mechanisms fostering
the emergence of cooperation in populations subjected to evolutionary
dynamics has attracted a lot of interest in ecology, biology, and
social sciences~\cite{nowak06,kollock98}. The problem has been tackled
through the formulation of simple games that neglect the microscopic
differences among distinct social and natural systems, thus providing
a general framework for the analysis of evolutionary
dynamics~\cite{maynard73,maynard82,gintis09}. Most of the classical
models studied within this framework made the simplifying assumption
that social systems are characterized by homogeneous structures, in
which the interaction probability is the same for any pair of agents and
constant over time~\cite{samuelson02}. However, the theory of complex networks has proven this assumption false
for real systems, by revealing that most natural and social networks exhibit large
heterogeneity and non--trivial interconnection
topologies~\cite{watts,strogatz,rev:albert,rev:newman}. It has
also been shown that the structure of a network has dramatic effects on the
dynamical processes taking place on it, so that complex networks
analysis has become a fundamental tool in epidemiology, computer
science, neuroscience, and social
sciences~\cite{rev:bocc,rev:doro2,rev:castellano}.

The study of evolutionary games on complex topologies has led to a
new way out for cooperation to survive in some paradigmatic cases such
as the Prisoner's Dilemma \cite{pachprl,pachpnas,prl,assenza} or the Public
Goods games \cite{pachnature,chaosPGG,percpggrev}. In particular, it has been
pointed out that the complex patterns of interactions among the agents
found in real social networks, such as scale-free distributions of the
number of contacts per individual or the presence of tightly-knit
social groups, tend to favor the emergence and persistence of
cooperation. This line of research, which brings together the tools
and methods from the statistical mechanics of complex networks and the
classical models of evolutionary game dynamics, has effectively became
a new discipline, known as evolutionary graph
theory~\cite{szaborev,jackson08,anxorev,percrev,grossrev}.

Recently, the availability of longitudinal spatio-temporal information
about human interactions and social relationships
\cite{eaglepentland,scott,isella,sthele_primary} has revealed that
social systems are not static objects at all: contacts among
individuals are usually volatile and fluctuate over
time~\cite{cattuto, saramaki}, face-to-face interactions are bursty
and intermittent~\cite{barabasi2005,bianconi2}, and agents motion exhibits
long spatio-temporal
correlations~\cite{gonzales,szell,sinatra}. Consequently, static
networks, constructed by aggregating in a single graph all the
interactions observed among a group of individuals across a given
period, can only be considered as simplified models of real networked
systems. For this reason, time-varying graphs have been 
introduced recently as a more realistic framework to encode time-dependent
relationships~\cite{kossinets,kostakos,tang_distance,holme_review, tang_sw}. In
particular, a time-varying graph is an ordered sequence of graphs
defined over a fixed number of nodes, where each graph in the sequence
aggregates all the edges observed within a certain
temporal interval.
The introduction of time as a new dimension of the graph gives rise to
a richer structure. Therefore, new metrics specifically designed to
characterize the temporal properties of graph sequences have been
proposed, and most of the classical metrics defined for static graphs
have been extended to the time-varying
case~\cite{nicosia_chapter,tang_sw,pan_paths,kovanen_motifs,tang_centrality,nicosia_components,mucha10}.
Recently, the study of dynamical processes taking place on time-evolving
graphs has shown that temporal correlations and contact recurrence
play a fundamental role in diverse settings such as random walks
dynamics~\cite{starnini_rw,perra_walking,ribeiro2013}, the spreading
of information and diseases~\cite{rocha_1,rocha_2,rocha_3}, and
synchronization~\cite{diazguilera}.

Here we study how the level of cooperation is affected when one
considers a more realistic picture, in which the interactions in a
social system are represented by time-varying graphs instead of
classical (static) ones. We consider a family of social dilemmas,
including the Hawk-Dove, the Stag Hunt, and the Prisoner's Dilemma
games, played by agents connected through a time-evolving topology
obtained from real traces of human interactions. We analyze the effect
of temporal resolution and correlations on the emergence of
cooperation in two paradigmatic data sets of human proximity, namely
the MIT Reality Mining~\cite{eaglepentland} and the
INFOCOM'06~\cite{scott} co-location traces.
We find that the level of cooperation achievable on time-varying
graphs depends crucially on the interplay between the speed at which
the network changes and the typical time scale at which agents update
their strategy. In particular, cooperation is facilitated when agents
keep playing the same strategy for longer intervals, while too
frequent strategy updates tend to favor defectors. Our results also
suggest that the presence of temporal correlations in the creation and
maintenance of interactions hinders cooperation, so that synthetic
time-varying networks in which link persistence is broken usually
exhibit a considerably higher level of cooperation. Finally, we show
that both the average size of the giant component and the weighted
temporal clustering calculated across different consecutive
time windows are indeed good predictors of the level of cooperation
attainable on time-varying graphs.

\begin{figure*}[t!]
\centering
\includegraphics[width=0.95\textwidth]{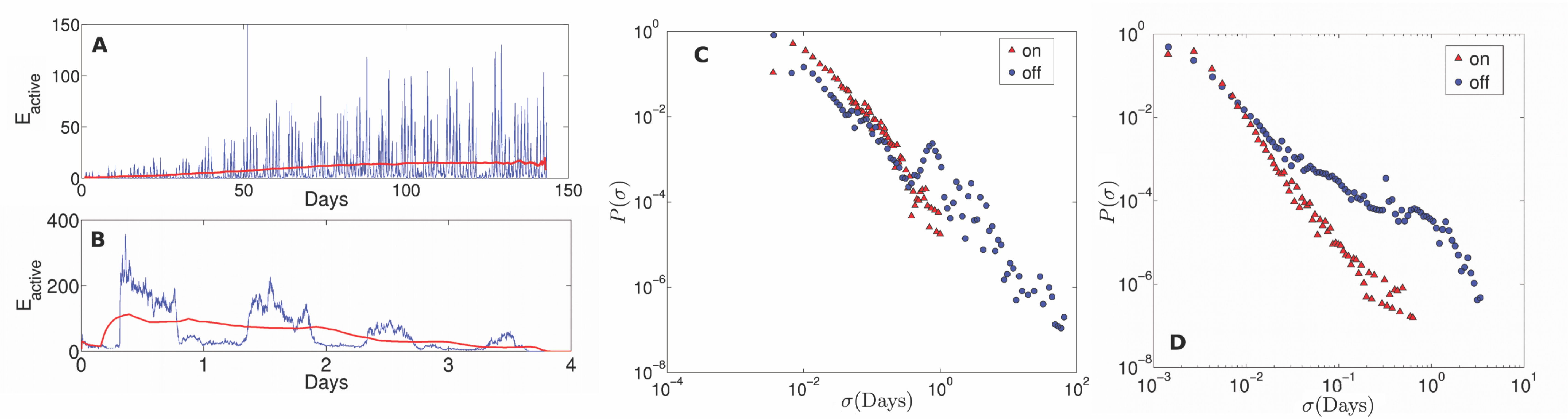}
\caption{(Color online). Activity patterns of human interactions. The number
  $E_{\text{active}}$ of links in the graph at time $t$ is reported as a
  function of time for MIT Reality Mining {\bf (a)} and
  INFOCOM'06 {\bf (b)}. Weekly and daily periodicities are
  visible. Red (light gray) lines display the moving averages, over a 1-month
  and a 1-day windows, respectively, revealing the non-stationarity of the
  sequences. Distributions of edge -active and -inactive
  periods (triangles and circles, respectively) for MIT Reality Mining
  {\bf (c)} and INFOCOM'06 {\bf (d)}. The data were log-binned. The
  peak at $\sigma \sim 1$ for the inactive periods corresponds to 24
  hours.}
\label{fig1}
\end{figure*}

\section{Evolutionary Dynamics on Time-Varying Graphs}
\subsection{Evolutionary dynamics of social dilemmas} 
%
%
We focus on the emergence of cooperation in systems whose
individuals face a social dilemma between two possible strategies:
{\em Cooperation} ($C$) and {\em Defection} ($D$). A large
class of social dilemmas can be formulated as in \cite{pachpnas} via a
two-parameter game described by the payoff matrix:
\begin{equation}
\bordermatrix{
  & C & D \cr
C & R & S \cr
D & T & P \cr} =
\bordermatrix{
  &  C & D \cr
C & 1 & S \cr
D & T & 0 \cr}\;,
\label{payoffs}
\end{equation}
where $R$, $S$, $T$, and $P$ represent the payoffs corresponding to the
various possible encounters between two players. Namely, when the two
players choose to cooperate, they both receive a payoff $R=1$ (for {\em
  Reward}), while if they both decide to defect they get $P=0$ (for
{\em Punishment}). When a cooperator faces a defector it gets the
payoff $S$ (for {\em Sucker}) while the defector gets $T$ (for {\em
  Temptation}).  In this version of the game, the payoffs $S$ and $T$
are the only two free parameters, and their respective
values induce an ordering of the four payoffs that determines the
type of social dilemma. We have in fact three different
scenarios. When $T>1$ and $S>0$, defecting against a cooperator
provides the largest payoff, and this corresponds to the {\em
  Hawk-Dove} game. For $T<1$ and $S<0$, cooperating with a defector is
the worst case, and we have the {\em Stag Hunt} game. Finally, for
$T>1$ and $S<0$, when a defector plays with a cooperator, we have at
the same time the largest (for the defector) and the smallest (for the
cooperator) payoffs, and the game corresponds to the {\em Prisoner's
  Dilemma}. In this work, we consider the three types of games by
exploring the parameter regions $T \in [0,2]$ and $S \in [-1,1]$.


In real social systems, each individual has more than one social
contact at the same time. This situation is usually
represented~\cite{anxorev} by associating each player $i,~
i=1,2,\ldots,N$ to a node of a {\em static} network, with adjacency
matrix $A=\{a_{ij}\}$, whose edges indicate pairs of individuals
playing the game. In this framework, a player $i$ selects a strategy,
plays a number of games equal to the number of her neighbors,
$k_i=\sum_{j}a_{ij}$, and accumulates the payoffs associated with each of
these interactions. Obviously, the outcome of playing with a neighbor
depends on the strategies selected by both players, according to the
payoff matrix in Eq.~\eqref{payoffs}. When all the individuals have
played with all their neighbors in the network, they update their
strategies as a result of an evolutionary process, i.e., according to
the total collected payoff. Namely, each individual $i$ compares her
cumulated payoff, $p_i$, with that of one of her neighbors, say $j$,
chosen at random. The probability $P_{i \rightarrow j}$ that agent $i$
adopts the strategy of her neighbor $j$ increases with the difference
$(p_j-p_i)$. Here we adopt the so-called \emph{Fermi update}
\cite{fermi1,fermi2} in which the probability that agent $i$ copies
the strategy of the randomly chosen neighbor $j$ reads:
\begin{equation}
P_{i \rightarrow j} = \dfrac{1}{1 + e^{-\beta (p_j - p_i) } } \;,
\label{eq:fermi-rule}
\end{equation}
where $\beta$ is a parameter controlling the smoothness of the
transition from $P_{i\rightarrow j}=0$ for small values of
$(p_j-p_i)$, to $P_{i\rightarrow j}=1$ for large values of
$(p_j-p_i)$. Notice that for $\beta\ll 1$ we obtain $P_{i\rightarrow
  j}\simeq 0.5$ regardless of the value of $(p_j-p_i)$, which
effectively corresponds to a random strategy update. On the other
hand, when $\beta\gg 1$ then $P_{i \rightarrow j}\simeq
\Theta(p_j-p_i)$, where $\Theta(x)$ is the Heaviside step function. Here 
we adopt $\beta=1$, although we have checked that the results are qualitatively similar 
for a broad range of values of $\beta$.

The games defined by the payoff matrix in Eq.~\eqref{payoffs}
and the use of a payoff-based strategy update rule have been thoroughly
investigated in static networks with different topologies. The main
result is that, when the network is fixed and agent strategies are
allowed to evolve over time, the level of cooperation increases with
the heterogeneity of the degree distribution of the network, with
scale-free networks being the most paradigmatic promoters of
cooperation~\cite{pachprl,pachpnas,prl}. However, in most cases
human contacts and social interactions are intrinsically dynamic and varying 
in time, a feature that has profound consequences on 
any process taking place over a social network. We explore here the role of time on the emergence of cooperation in time-varying
networks.

\begin{figure*}
\centering
\includegraphics[width=6.8in]{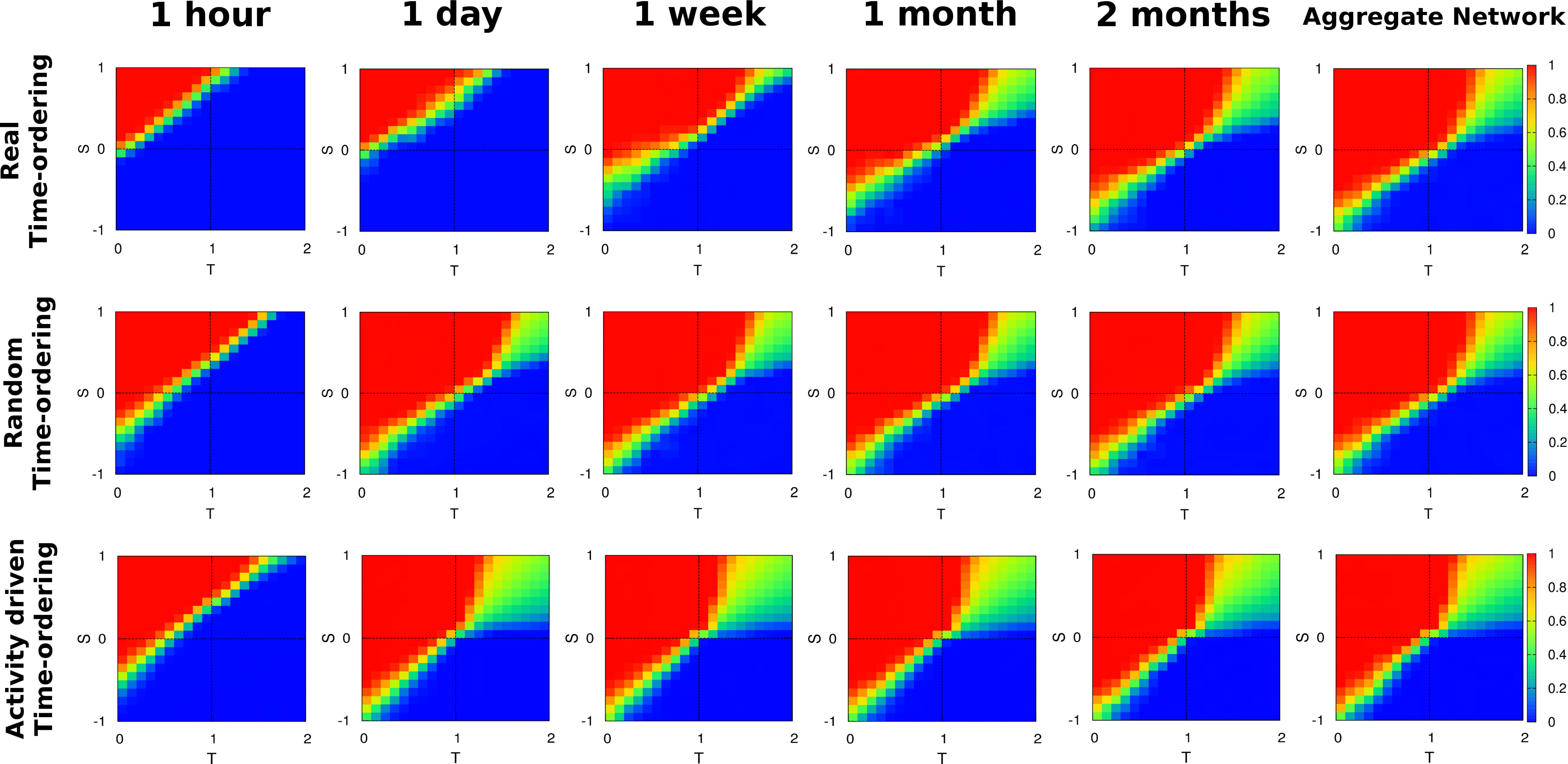}
\caption{(Color online). Cooperation diagrams for the MIT Reality
  Mining data set. Fraction of cooperators at the equilibrium as a
  function of the temptation to defect (T) and of the sucker's score
  (S) for different values of the interval $\Delta t$ between two
  successive strategy updates. From left to right, the diagrams
  correspond to $\Delta t$ equal to 1 hour, 1 day, 1 week, 1 month, 2
  months, and to the entire observation period $M\tau\simeq 5$
  months. The diagrams in the top row correspond to time-varying
  graphs with original time ordering, those in the middle row are
  obtained for the same values of $\Delta t$ but on randomized
  time-varying graphs, while the bottom row reports the results
  obtained on synthetic networks constructed through the
  activity-driven model. The results are averaged over 50 different
  realizations. Red (light gray) corresponds to $100\%$ of cooperators while blue (dark gray) 
  indicates $100\%$ defectors. By focusing on a row, and
    proceeding from left to right, it is evident that there exists a
    value of the update interval $\Delta t$ above which the
    differences in the cooperation diagram are mostly limited to the
    region which separates the two phases (100\% defectors and 100\%
    cooperators), while the rest of the phase diagram is already
    indistinguishable from that corresponding to the aggregate
    graph. Moreover, for a fixed value of $\Delta t$, we observe that
    randomized and synthetic sequences are associated with an overall
    larger level of cooperation than the original ones. See
    Fig.~\ref{fig:figs2} for additional details.}
\label{fig2}
\end{figure*}

\subsection{Temporal patterns of social interactions} 
In the following, we consider two data sets describing the temporal patterns of human
interactions at two different time scales. The first data set has been
collected during the MIT Reality Mining
experiment~\cite{eaglepentland}, and it includes information about
the spatial proximity of a group of students, staff, and faculty members
at the Massachusetts Institute of Technology, over a period of six
months.  The resulting time-dependent network has $N=100$ nodes and
consists of a time-ordered sequence $\{G_1, G_2, \ldots, G_M\}$ of
$M=41291$ graphs (snapshots), each graph representing proximity
interactions during a time interval of $\tau=5$ minutes. Remember that
each graph $G_m$ ($m=1,\ldots,M$) accounts for all the instantaneous
interactions taking place in the temporal interval $[(m-1)\tau, m
  \tau]$.  The second data set describes co-location patterns, over a
period of four days, among the participants of the INFOCOM'06
conference~\cite{scott}. In this case, the resulting time-dependent
network has $N=78$ nodes, and it contains a sequence of $M=2880$ graphs
obtained by registering users co-location every $\tau=2$ minutes.
Additional details about the two data sets are reported in Appendix A. 

The frequency of social contacts is illustrated in Fig.~\ref{fig1} [panels
(a) and (b)], where we report the number of active links at time $t$,
$E_{\text{active}}$, as a function of time. In the MIT Reality Mining data
set, social activity exhibits daily and weekly periodicities,
respectively due to home--work and working days-weekends cycles. In
addition to these rhythms, we notice a non-stationary behavior which
is clearly visible when we plot the activity averaged over a 1-month
moving window [red line in panel (a)]. In the INFOCOM'06 data set we
observe a daily periodicity and a non-stationary trend which is due,
in this case, to a decreasing social activity in the last days of the
conference as seen by aggregating activity over 24 hours [red line in
panel (b)].
We also report in Fig.~\ref{fig1} [panels (c) and (d)] the distributions
$P(\sigma)$ of contact duration, $\sigma \equiv \sigma_{\text{on}}$, and of
inter-contact time, $\sigma \equiv \sigma_{\text{off}}$ (i.e., the interval
between two consecutive appearances of an edge). As it is often the
case for human dynamics~\cite{barabasi2005}, the distributions of
contact duration and inter-contact time are heterogeneous. For the MIT
data set, an active edge can persist up to an entire day, while
inactive intervals can last over multiple days and weeks; similar
patterns are observed in the INFOCOM'06 data set, where some edges
remain active up to one entire day and inter-contact times span almost
the whole observation interval.
Edge activity exhibits significant correlations over long periods of
time. In particular, the autocorrelation function of the time series
of edge activity shows a slow decay, up to lags of $6-8$ hours for
the MIT data set, and of $3-4$ hours for INFOCOM'06, after which the
daily periodicity becomes dominant (as displayed in Fig. \ref{fig:figs1}). 

\begin{figure*}[!t]
\centering
\includegraphics[width=7.0in]{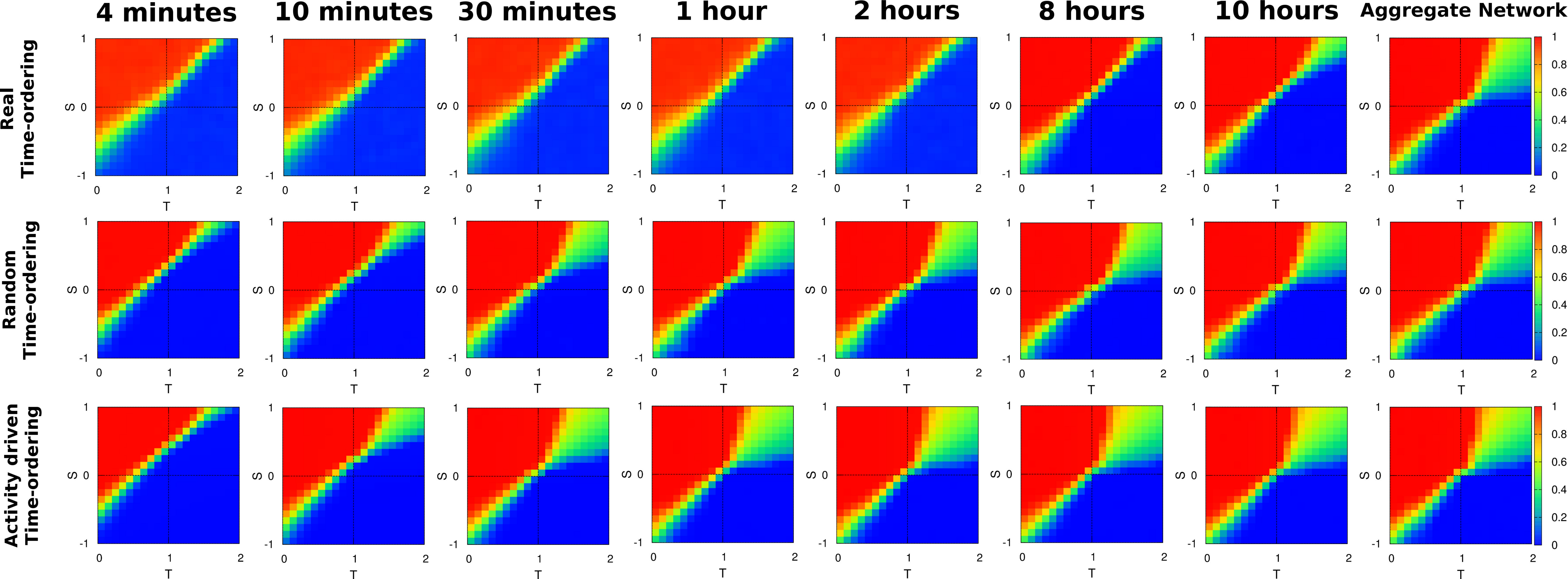}
\caption{(Color online). Cooperation diagrams for the INFOCOM data
  set.  Fraction of cooperators at the equilibrium as a function of
  the temptation to defect (T) and of the sucker's score (S) for
  different values of the interval $\Delta t$ between two successive
  strategy updates. From left to right, the diagrams correspond to
  $\Delta t$ equal to 4 minutes, 10 minutes, 30 minutes, 1 hour, 2
  hours, 8 hours, 10 hours, and $M\tau\simeq 4$ days. The top, middle,
  and bottom row report, respectively, the results for the original
  data set, the reshuffled time-varying graph, and synthetic graphs
  constructed through the activity-driven model. The results are
  averaged over 50 different realizations. Red (light gray) corresponds to $100\%$
  of cooperators, while blue (dark gray) indicates $100\%$ defectors. As in Fig. \ref{fig2}, the comparison of the diagrams corresponding to the same update interval shows that differences appear in the region that separates the full cooperation and full defection phases and that these differences favor (in terms of degree of cooperation) the randomized and synthetic sequences. See Fig.~\ref{fig:figs3} for additional details.}
\label{fig3}
\end{figure*}

\section{Evolution of Cooperation in Time-varying Networks}

\subsection{Cooperation Diagrams}

To simulate the game on a time-varying topology
$\{G_m\}_{m=1,\ldots,M}$, we start from a random distribution of
strategies, so that each individual initially behaves either as a
cooperator or as a defector, with equal probability.  The simulation
proceeds in \textit{rounds}, where each round consists of a
\textit{playing} stage followed by a \textit{strategy update}. In the
first stage, each agent plays with all her neighbors on the first
graph of the sequence, namely on $G_1$, and accumulates the payoff
according to the matrix in Eq.~\eqref{payoffs}. Then the graph changes,
and the agents employ the same strategies to play with all their
neighbors in the second graph of the sequence, $G_2$. The new payoffs
are summed to those obtained in the previous iteration. The same
procedure is then repeated $n$ times with $n$ such that $n\tau$
is equal to a chosen interval $\Delta t$, which is the \textit{strategy update
interval}. At this point, the playing stage terminates and agent
strategies are updated following the {\em Fermi update} in Eq. (\ref{eq:fermi-rule}).
After the agents have updated their strategy, their payoff is
reset to $0$ and they start another round, during the subsequent time
interval of length $\Delta t=n\tau$, as described above.

To evaluate the degree of cooperation obtained for a given value of
the strategy update interval $\Delta t$ and a pair of values $(T,S)$,
we compute the average fraction of cooperators $\langle C(T,S)_{\Delta
  t} \rangle$:
\begin{equation}
  \langle C(T,S)_{\Delta t} \rangle = \dfrac{1}{Q} \sum_{i=1}^{Q}
  \dfrac{N_c^i}{N}\,,
\label{cmed}
\end{equation}
where $N_c^i$ is the number of cooperators found at time $i\,\Delta t$
and $Q$ is the total number of rounds played. In general, we set $Q$
large enough to guarantee that the system reaches a stationary state in which the level of cooperation remains roughly constant.

We have simulated the system using different values of $\Delta
t$. Notice that for smaller value of $\Delta t$, the time scale of
the strategy update is comparable with that of the graph evolution, while when
$\Delta t$ is equal to the entire observation period $ M \tau$ the
game is effectively played on a static topology, namely the weighted
aggregated graph corresponding to the whole observation interval.
We focus now on the top panels of Figs.~\ref{fig2} and \ref{fig3}, where we show how the average fraction of cooperators
depends on the parameters $S$ and $T$ and on the length $\Delta t$ of
the strategy update interval. We considered six values of $\Delta t$
for the MIT data set, from $\Delta t=1$ hour up to the
whole observation interval, and eight values for INFOCOM'06, ranging
from minutes up to the aggregate network.

At first glance, we notice that the rightmost diagrams in both
figures, which correspond to $\Delta t=M\tau$, are in perfect
agreement with the results of evolutionary games played on static
topologies reported in the literature (see, {\em
  e.g.},~\cite{pachpnas,anxorev}).  If we look at the
  cooperation diagrams obtained by increasing the value of $\Delta t$
  in the original sequences of graphs (top panels of Figs.~\ref{fig2}
  and \ref{fig3}), we notice an increase of the area of the red
  region, which corresponds to configurations in which 100\% of the
  nodes are cooperators at the stationary state.  In particular, for
  MIT Reality Mining (Fig.~\ref{fig2}), the fraction of cooperators
  increases up until $\Delta t=2$ months, after which the cooperation
  diagram is practically indistinguishable from that obtained on the
  static aggregated graph.

As we pointed out above, edge activation patterns show non-trivial
correlations.  To highlight the effects of temporal correlations and
of periodicity in the appearance of links in the real data sets, we
have simulated the games also on randomized time-varying graphs and on
synthetic networks generated through the activity-driven
model~\cite{Perra2012} (See Appendix B for details on the activity-driven model). 
The results for randomized graphs and
activity-driven graphs are reported, respectively, in the middle and
in bottom panels of Figs.~\ref{fig2} and \ref{fig3}.

Randomized time-varying graphs are obtained by uniformly reshuffling
the original sequences of snapshots. In this case, the frequency of
each pairwise contact is preserved equal to that of the original data
set. However, the temporal correlations of these contacts, namely the
persistence of an edge during consecutive time snapshots, are
completely wiped out.  As expected, for $\Delta t=M\tau$ the
cooperation diagrams obtained on the reshuffled sequences (middle
rightmost panels of Figs.~\ref{fig2} and \ref{fig3}) are identical
to those obtained on the corresponding original data sets (top
rightmost panels).  In fact, when $\Delta t=M\tau$ each agent plays
with all the contacts she has seen in the whole observation interval,
with the corresponding weights, before updating her strategy, and thus
the frequencies of contacts are the only ingredients responsible for
the emergence of cooperation.
Conversely, for smaller values of $\Delta t$, the importance of the
temporal correlations of each pairwise contact becomes clear since the
cooperation diagrams for randomized and original networks are very
different in both data sets. In fact, for the randomized graphs, the
cooperation levels at $\Delta t=1$ week and $\Delta t=2$ hours for the
Reality and INFOCOM data sets, respectively, are comparable to those for
$\Delta t=M\tau$. This points out that cooperation is enhanced by
destroying the temporal correlations of pairwise contacts.

Little differences are observed between activity-driven synthetic
networks and the corresponding graph sequence randomizations (results
shown in the bottom panels of Figs.~\ref{fig2} and \ref{fig3}). In
this case not only are temporal correlations wiped out, but also the
microscopic structure of each snapshot is replaced by a graph having a
similar density of links. This rewiring distributes links more
heterogeneously than in the original and the randomized sequences (see
Appendix \ref{activity} for details). The cooperation diagrams of
activity-driven networks show a further increase of the cooperation
levels for even smaller values of the strategy update interval,
$\Delta t$, than in the case of Random graphs. Namely, for $\Delta
t=1$ day in Reality Mining (Fig.~\ref{fig2}) and for $\Delta t=30$
minutes in INFOCOM (Fig.~\ref{fig3}), we already recover the
cooperation levels of $\Delta t=M\tau$. These results
  indicate that defectors take advantage of the volatility of edges,
  and that cooperation emerges only when the interval between two
  consecutive strategy updates is large enough. A more detailed
  visualization of the differences between the phase diagrams obtained
  in the original sequence for different values of $\Delta t$ and
  those observed for the randomized and synthetic networks is reported
  in Figs.~\ref{fig:figs2}  and \ref{fig:figs3}.

\subsection{Structural analysis of time varying networks}
\label{secd}

The reported results suggest that the ordering, persistence, and 
distribution of edges over consecutive time windows are all fundamental 
ingredients for the success of cooperation. In general, a small value of $\Delta t$ 
in the original data sets corresponds to playing the game on a sparse graph, 
possibly comprising a number of small components, in which nodes are 
connected to a small neighborhood that persists rather unaltered over consecutive 
time windows. The small size of the isolated clusters and the persistence of the connections within them allow defectors to spread their strategy efficiently. In the following, we will test 
this hypothesis by characterizing the structure of the original and randomized versions of the
time-varying graphs.

\begin{figure}[!t]
\centering

\begin{center}
  \includegraphics[width=.49\textwidth]{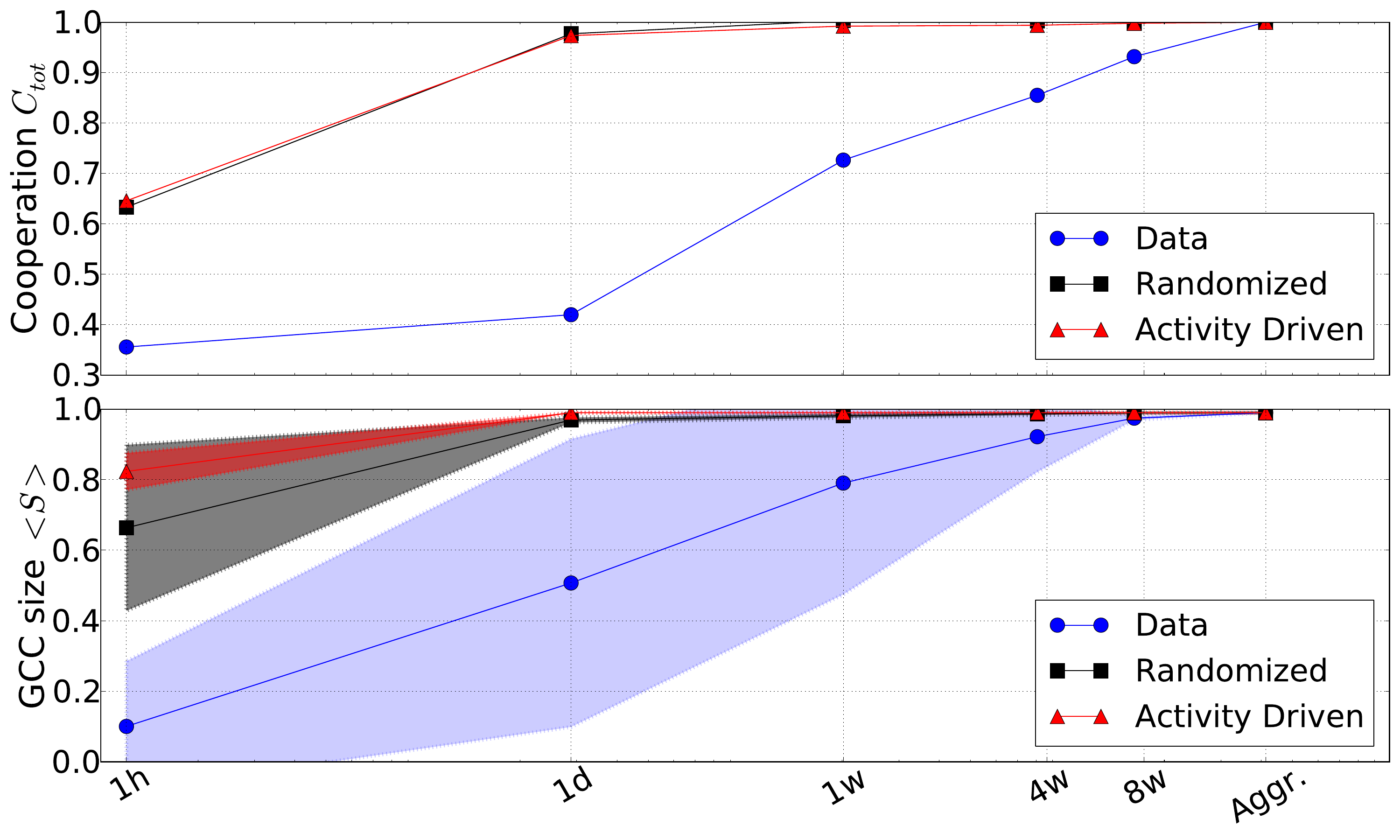}\\
  \includegraphics[width=.49\textwidth]{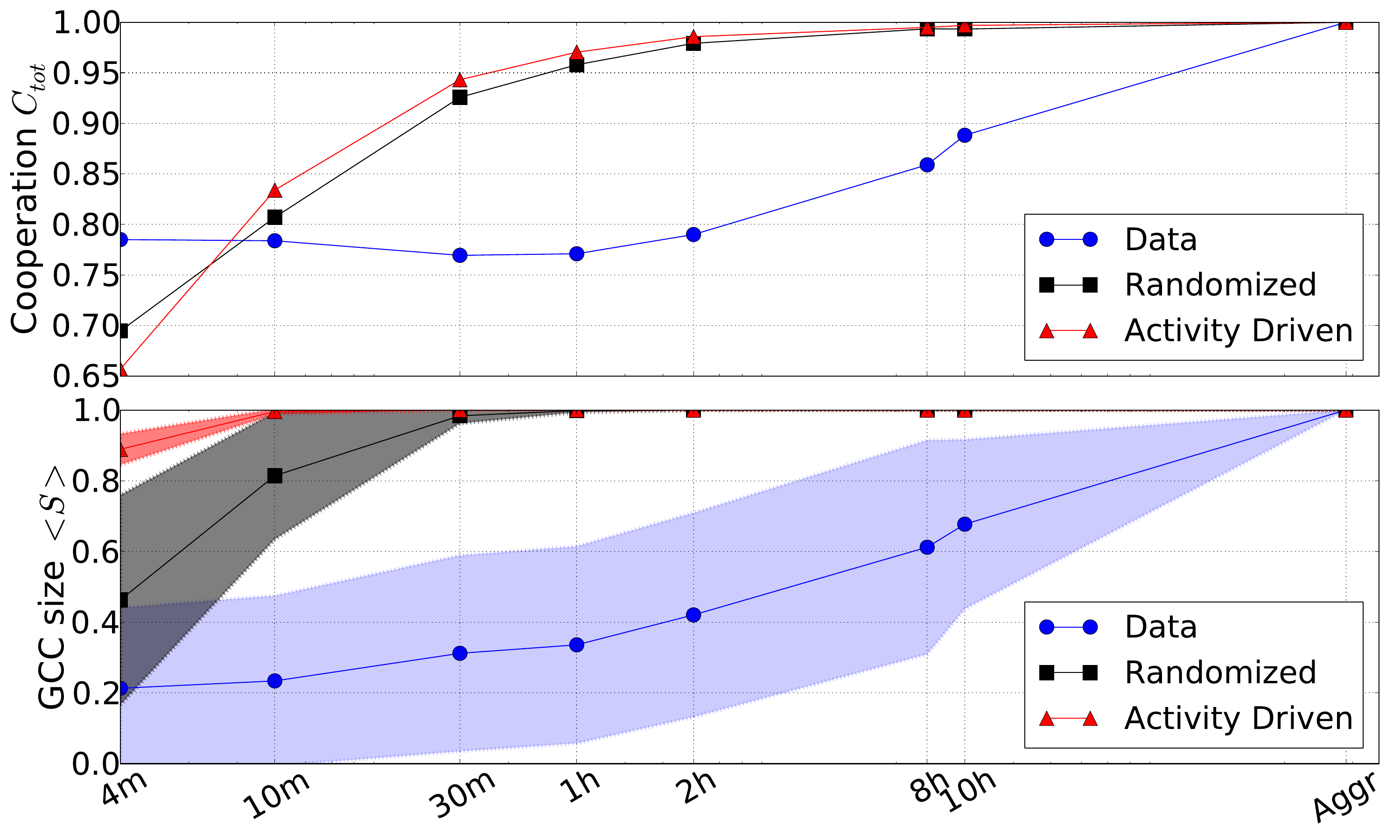}
\end{center}
\caption{\label{fig:giant}(Color online). Cooperation level and size of the Giant
    Component. Overall cooperation level $C_{tot}(\Delta t)$ and
  average size of the giant component $\avg{\mathcal{S}}$ as a
  function of the aggregation interval $\Delta t$ for MIT Reality
  Mining (top panels) and INFOCOM'06 (bottom panels). Blue circles
  correspond to the original data, black squares to the reshuffled
  networks and red triangles to the activity-driven model. The shades
  indicate the standard deviation of $\avg{\mathcal{S}}$ across the
  sequence of graphs for each value of $\Delta t$. Notice that the
  typical size of the giant component at time-scale $\Delta t$
  correlates quite well with the observed cooperation level at the
  same time-scale.}
\end{figure}

In order to investigate the dependence of cooperation on the strategy
update interval $\Delta t$, we computed the average fraction
$\avg{\mathcal{S}}$ of nodes in the giant component of the
graphs as a function of $\Delta t$ for the original data sets and for
the reshuffled and synthetic sequences of snapshots. The results reported in Fig.~\ref{fig:giant} indicate that
for a given value of $\Delta t$, the
giant component of graphs in the randomized sequences or in
the activity-driven model is larger than that of graphs in the
original ordering. The lack of temporal correlations between
consecutive time snapshots in randomized and activity-driven networks
produces an increase in the number of ties between different agents of
the population even for small values of $\Delta t$. In addition, the
more homogeneous distribution of links within the snapshots of the
activity-driven network further increases the mixing of the agents and
thus enlarges the size of the giant connected component compared to that of
randomized graphs.

In Fig.~\ref{fig:giant}, we also show the overall level of
cooperation observed at a given aggregation scale $\Delta t$, $C_{\text{tot}}(\Delta
t)$, defined as:
\begin{equation*}
  C_{\text{tot}}(\Delta t) =
  \frac{1}{C_{\text{tot}}(M\tau)}\int_{0}^{2}\!\!\!\mathrm{d}T\!\int_{-1}^{1}\!\!\!
  C(T,S)\,\mathrm{d}S\,.
\end{equation*}
Notice that $C_{\text{tot}}(\Delta t)$ is divided by the value
$C_{\text{tot}}(M\tau)$ corresponding to the whole observation interval, so
that $C_{\text{tot}}\in[0,1]$. The value of $\Delta t$ at which
$\avg{\mathcal{S}}$ is comparable with the number of nodes $N$,
i.e. when $\avg{\mathcal{S}}\simeq 1$, coincides with the value of
$\Delta t$ at which the cooperation diagram becomes indistinguishable
from that obtained for the aggregate network, $C_{\text{tot}}(\Delta t)\simeq
1$, for both the original and the reshuffled sequences of
snapshots. This result confirms that the size of the giant connected
component of the graph corresponding to a given aggregation interval
plays a central role in determining the level of cooperation
sustainable by the system, in agreement with the experiments discussed
in~\cite{wang_scirep_2012} for the case of static complex networks.%

\begin{figure}[!t]
  \begin{center}
    \includegraphics[width=0.65\columnwidth]{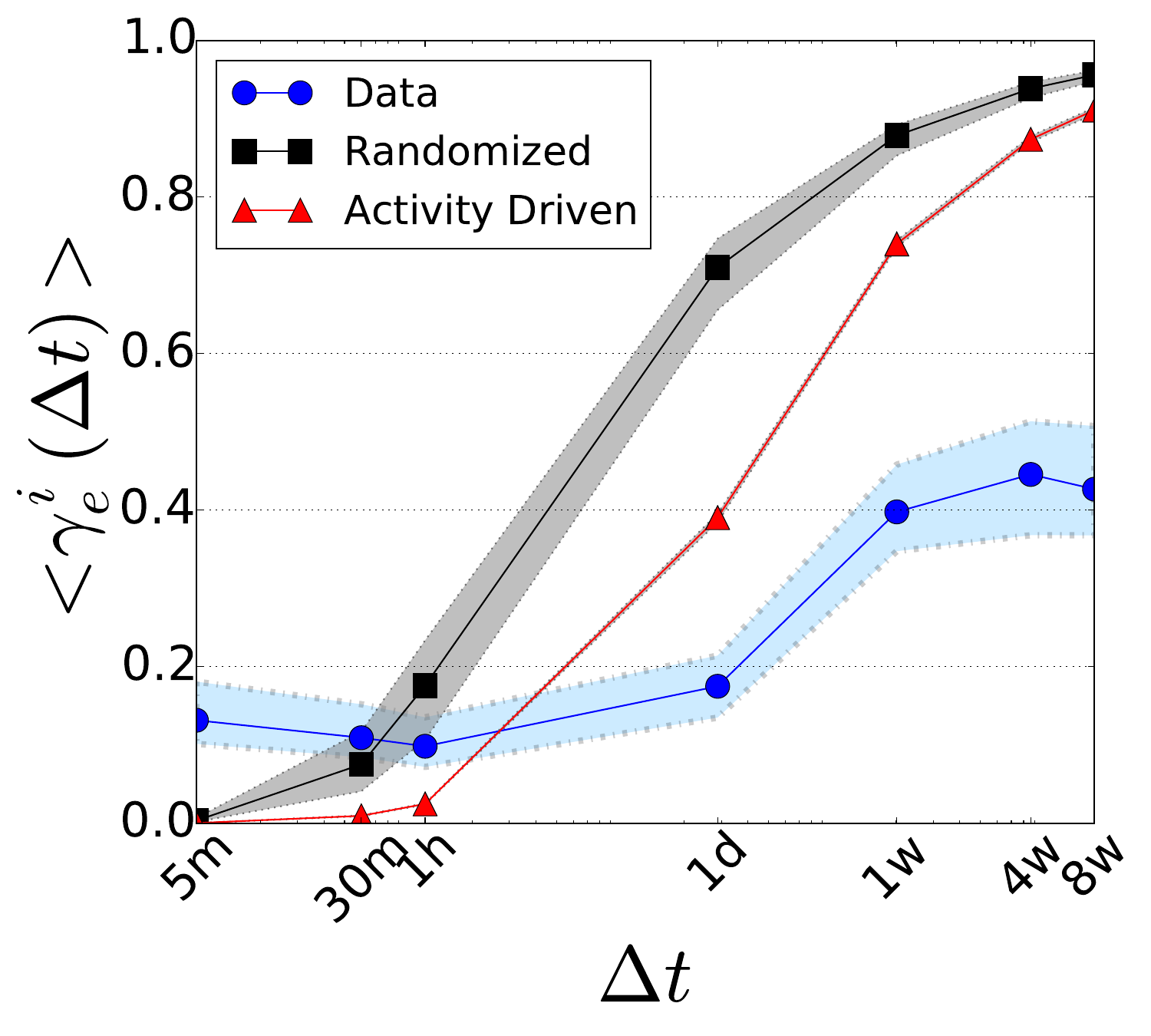}\\
    \includegraphics[width=0.65\columnwidth]{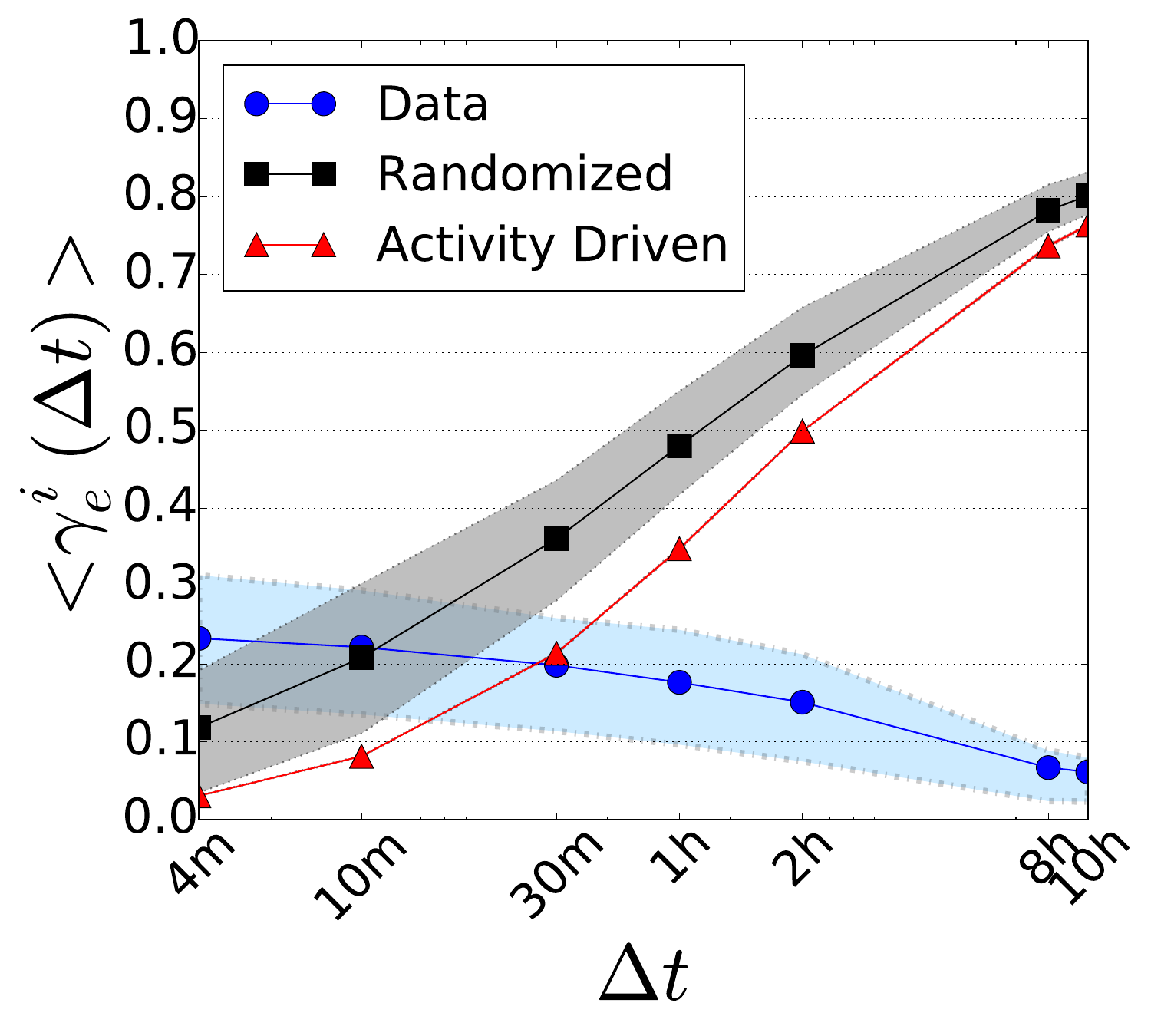}\\
  \end{center}
  \caption{(Color online). Extremal temporal clustering $\gamma_e^i$ as a function of
    the strategy update interval $\Delta t$ on real data sets (blue
    dots). Top (bottom) panel refers to Reality Mining
    (INFOCOM). Black squares correspond to randomly reshuffled
    sequences and red triangles to activity-driven synthetic
    networks. In both data sets we notice that, for small values of
    $\Delta t$, real data display an higher value of clustering
    (persistence) than synthetic cases followed by a transition value
    of $\Delta t$ above which we observe a rapid increase in the
    clustering of synthetic cases such that the previous situation is
    inverted.}
  \label{fig5}
\end{figure}

We also investigate the role of edge correlations between consecutive graphs on
the observed cooperation level.
To this aim, we analyze the temporal
clustering $\gamma_e^i$ (see Appendix \ref{clustering}), which captures the average
tendency of edges to persist over time. In Fig.~\ref{fig5} we plot the
evolution of the temporal clustering as a function of the strategy
update interval $\Delta t$. The results clearly reveal that, for small
values $\Delta t$, the persistence of ties in the two original data
sets is larger than in the randomized and the activity-driven graphs.
Overall, for small $\Delta t$, the large temporal clustering and the small average size of the giant component indicate that the graphs are composed of small clusters of nodes whose composition changes
very slowly compared to the faster mixing observed in the randomized sequences.
Thus, these are two ingredients hindering the cooperation levels in 
the original data sets: the size of the giant 
component, and the internal arrangement of connections within the different 
components.

\begin{figure}[!t]
\centering  
  \includegraphics[width=0.99\columnwidth]{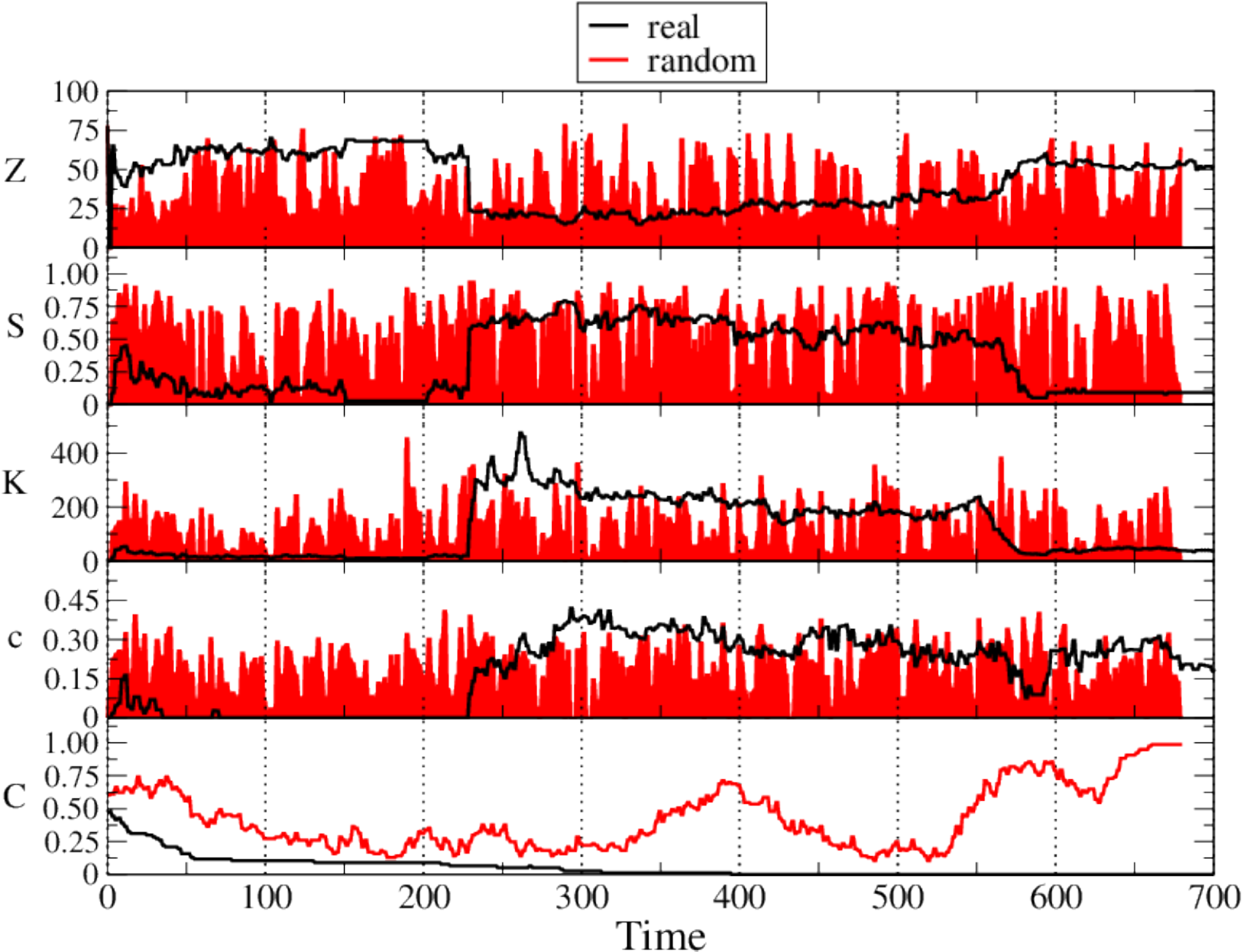}
  \caption{(Color online). Evolution of cooperation and topological quantities as a function of time $t$ for the INFOCOM data set in the Harmony Game ($T=0.9, S=0.2$). From top to bottom we display the number of components $Z$, fraction of nodes in the giant component $S$, number of links $K$, topological clustering coefficient $c$, and fraction of cooperators $C$ as a function of time. Black (red or gray in b/w) line corresponds to the original (randomized) data set.}
  \label{fig6}
\end{figure}

As an example of the negative effect on cooperation of the
combination of the two latter ingredients, we consider a pair of $\left( T, S\right)$ values in the Harmony Game (HG) regime ($0<T<1$ and
$0<S<1$). The evolutionary dynamics of the HG in the well mixed
(all-to-all) regime drives the system towards full cooperation.
Even though this regime is the best scenario for the promotion of
cooperation, the real time varying graphs exhibit small cooperation
levels for small $\Delta t$.
This is mainly due to the high level of segregation of interactions in disconnected and small clusters. 
Under these conditions (far from the well mixed hypothesis), the HG behaves differently in the regimes $T<S$ and $S<T$ (these two regimes are separated by the solid line $S=T$ in the panels of Figs. \ref{fig2} and \ref{fig3}). While in the first regime, $T<S$, the pairwise encounters between a cooperator and a defector yield more benefit to the former, this is not the case when $S<T$. Thus, when the population is segregated into small (and persistent) clusters containing a small number of nodes, defection easily prevails in the region $S<T$ of the HG. This counterintuitive result is obtained for the original time varying graphs (especially for that of MIT) and small $\Delta t$. The time evolution of the interaction patterns of the real data sets confirms the structural roots of this behavior.

In Fig.~\ref{fig6} we display, for the case of INFOCOM, the number of
components $Z$, the number of links $K$, the size of the giant component $S$, and the topological clustering coefficient $c$ ({\em i.e.} the probability that two neighbors of a given node are also connected) as a function of time $t$.
Similar results are obtained for the MIT data set. We notice considerable differences between the time
evolution of these quantities in the original time series and in the
randomized ones. Namely, in the original time-varying graph we observe
a long initial time window during which the real network displays a
large number of components and poor
connectivity. This period is then followed by another long period
characterized by the appearance of a connected and clustered giant
component.  As expected, the randomized graph does not show this
persistent behavior. In the bottom panel, we show the evolution of
cooperation when the update interval is set to its minimum value,
$\Delta t=4$ minutes. As can be observed, starting at time $0$
would imply that the initial fraction of cooperators face a rather
complicated scenario for their survival even in the HG regime.

Turning our attention back to Fig. \ref{fig5}, we notice that, as $\Delta t$ increases, the link persistence grows similarly in the randomized and activity-driven networks.
%
This growth points out that the randomization of snapshots in one null model and the redistribution of links in the other one make the ties more stable as $\Delta t$ increases. 
This stabilization, however, does not lead to a decrease in cooperation since it is combined with the fast increase with $\Delta t$ of the size of the giant component.

%
%

%

\section{Conclusions}

Although the impact of network topology on the onset and
persistence of cooperation has been extensively studied in recent 
years, the recent availability of data sets with time-resolved
information about social interactions allows for a deeper investigation of
the impact of time-evolving social
structures on evolutionary dynamics. 
Here we addressed two crucial questions: does the
interplay between the time scale associated with graph evolution and that corresponding to the strategy update affect the classical results about the enhancement of cooperation driven by
network reciprocity? And what is the role of the time correlations of temporal networks in the evolution of cooperation? 
The importance of the competition between the time scale of social ties and their corresponding outcome (here the games played and the benefits obtained) and the update of strategies have been recently addressed \cite{cuesta,pachtraul}. However, in our work we attempted to go one step further by relying on two empirical data sets incorporating the two ingredients whose impact over the evolution of cooperation we want to evaluate: the time scale of social interactions and their temporal correlations. 

Our results confirm that, for all four social dilemmas studied in this work, cooperation is seriously hindered when (i) agent strategy is updated too frequently with respect to the typical time scale of agent interaction, and (ii) realistic link temporal correlations are present. 
This phenomenon is a consequence of the relatively small size of the giant component of the graphs obtained at small aggregation intervals. However, when the temporal sequence of social contacts is replaced by randomized or synthetic time-varying networks preserving the original activity attributes of links or nodes but breaking the original temporal correlations, the structural patterns of the network at a given time scale of strategy update changes dramatically from those observed in real data. As a consequence, the effects of temporal resolution over cooperation are smoothed and, by breaking  the real temporal correlations of social contacts, cooperation can emerge and persist even for moderately small time periods between consecutive strategy updates.


Our findings suggest that the frequency at which the connectivity of a given system is sampled has to be carefully chosen, according with the typical time scale of the social interaction dynamics.  For
instance, as stock brokers might decide to change strategy after just a couple of interactions, other processes such as trust formation in business or collaboration networks are likely to be better described
as the result of multiple subsequent interactions. These conclusions are also supported by the results of a recent paper by Ribeiro {\em et al.} \cite{ribeiro2013} in which the effects of temporal aggregation interval $\Delta t$ in the behavior of random walks are studied. 
One limitation of the current work comes from the fact that the used data sets have not been specifically collected in order to study cooperation spreading on networks and might represent therefore a suboptimal network substrate for the dynamical process under study. 
At the same time, these empirical data sets arguably contain the most direct measurement of human interaction upon which any social interaction mechanism is then built up and --already at this simple level of face-to-face interaction-- contain rich and non-trivial structures and phenomena.
One example of this is the fundamental role played by the real-data time correlations in dynamical processes on the graph, which calls for more models of temporal networks and for a better understanding of their nature. 
In a nutshell, our results point out that one should always bear in mind
that both the over- and the under-sampling of time-evolving social
graph and the use of the finest/coarsest temporal resolution could
substantially bias the results of a game-theoretic model defined on the
corresponding network. These results pave the way to a more detailed
investigation of social dilemmas in systems where both structural
and temporal correlations are incorporated in the interaction
maps.

\begin{acknowledgments}
This work was supported by the EU LASAGNE Project, Contract No.318132
(STREP), by the EU MULTIPLEX Project, Contract No.317532 (STREP), by the EU
PLEXMATH Project, Contract No. 317614 (STREP), by
the Spanish MINECO under projects MTM2009-13848 and FIS2011-25167
(co-financed by FEDER funds), by the Comunidad de Arag\'on (Grupo
FENOL) and by the Italian TO61 INFN project. J.G.G. is supported by
Spanish MINECO through the Ram\'on y Cajal program. G.P. is supported
by the FET project ``TOPDRIM" (IST-318121). R.S. is supported by a James S. McDonnell Foundation Postdoctoral Fellowship.
V.L. was supported by the EPSRC project GALE EP/K020633/1
\end{acknowledgments}

\appendix

\section{Data sets Description}

\begin{figure}[ht!]
  \begin{center}
    \includegraphics[width=.9\columnwidth]{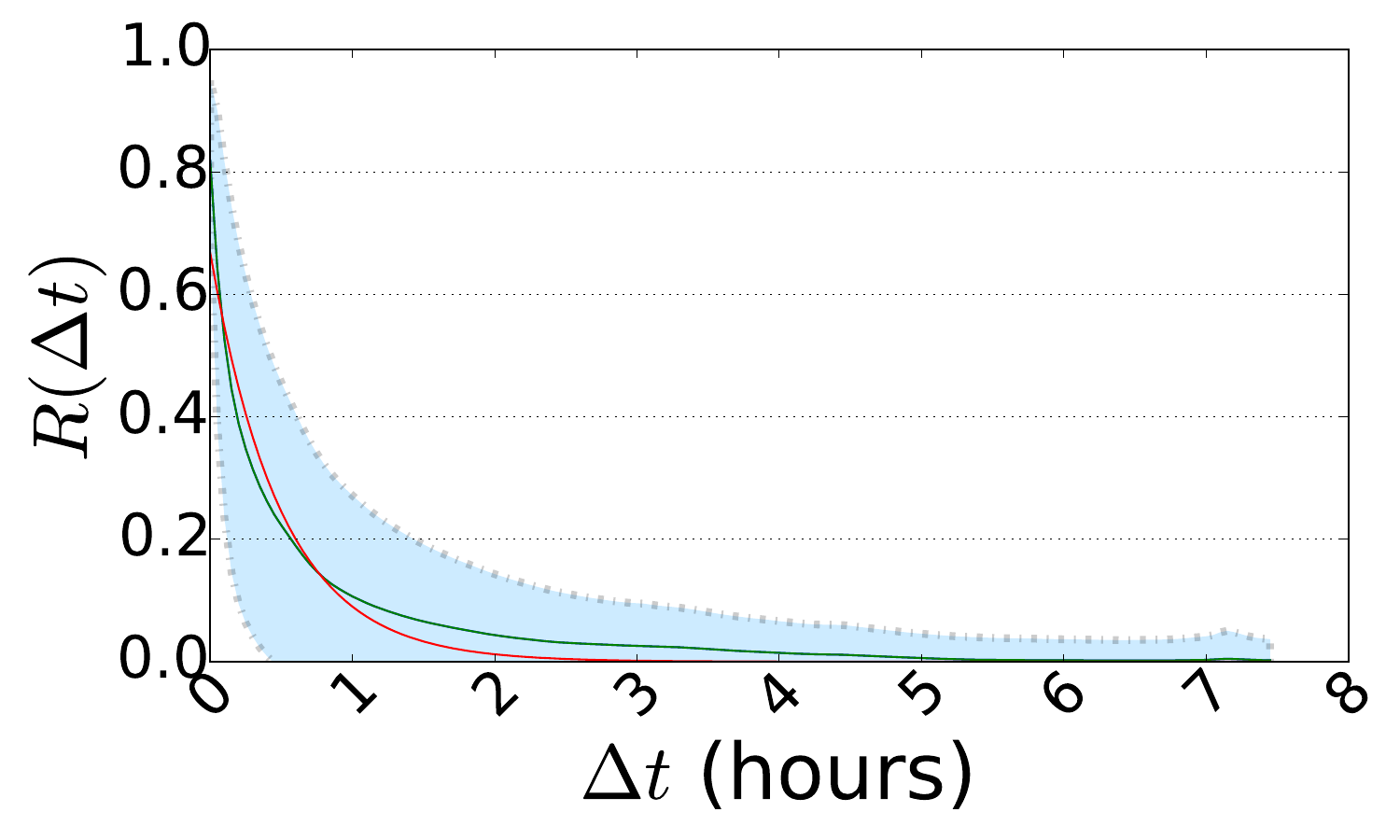}\\
    \includegraphics[width=.9\columnwidth, height=.58\columnwidth]{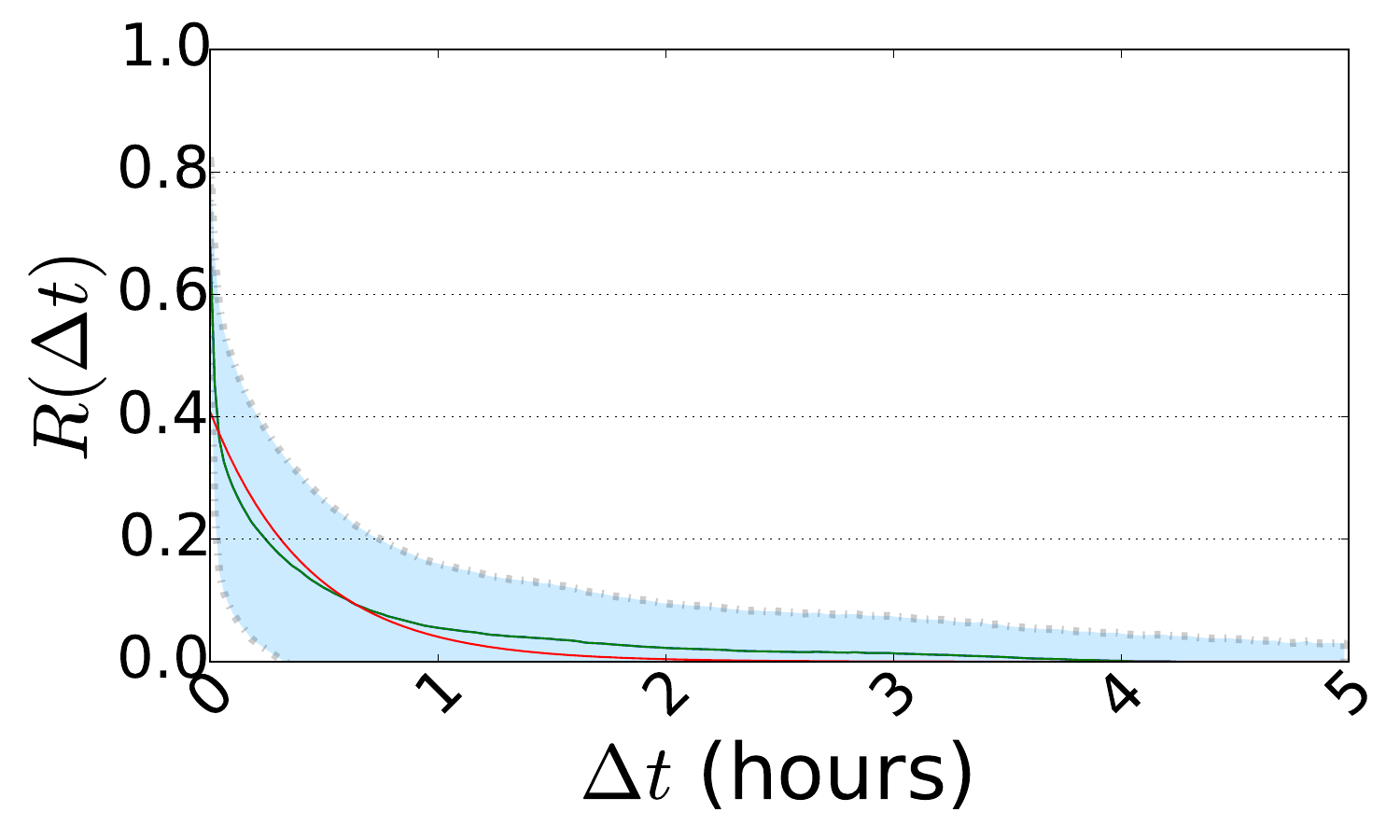}
  \end{center}
  \caption{(Color online). Autocorrelation fucntion $R$ for edge activations, averaged over all the active edges in the data sets, Reality ({\it top}) and INFOCOM ({\it bottom}). The long lasting correlations in the edge activity of both data sets (up to 6-8 hours for Reality and 3-4 hours for INFOCOM) represent one of the main ingredients behind the different cooperation outcomes. Red (gray) line corresponds to the exponential fit of the data.}
  \label{fig:figs1}
\end{figure}

In the following, we introduce the principal characteristics of the two data sets used in our study. As stated in the main text, one of the reasons behind the emergence of cooperation is the persistence of interactions. A way to gauge such persistence is measuring the autocorrelation function $R$ of our time series as shown in Fig.~\ref{fig:figs1}.

\subsection{MIT Reality Mining data set}

The data set describes proximity interactions collected through the
use of Bluetooth-enabled phones~\cite{eaglepentland}. The phones were
distributed to a group of 100 users, composed by 75 MIT Media
Laboratory students and 25 faculty members. Each device had a unique
tag and was able to detect the presence and identity of other devices
within a range of 5--10 meters.  The interactions, intended as
proximity of devices, were recorded over a period of about six
months. In addition to the interaction data, the original data set
included also information regarding call logs, other Bluetooth devices
within detection range, the cell tower to which the phones were
connected, and information about phone usage and status. Here, we
consider only the contact network data, ignoring any other contextual
metadata.
The resulting time-varying network is an ordered sequence of 41291
graphs, each having N=100 nodes. Each graph corresponds to a proximity scan
taken every 5 minutes. An edge between two nodes indicates that the
two corresponding devices were within detection range of each other
during that interval. We refer to such links as {\it active}. During
the entire recorded period, 2114 different edges have been detected as
active, at least once.  This corresponds to the aggregate graph having
a large average node degree $\avg{k}\simeq 42$. However, this is an
artefact of the aggregation; the single snapshots tend to be very
sparse, usually containing between 100 and 200 active edges.
 
\subsection{INFOCOM'06 data set}
The data set consists of proximity measurements collected during the
IEEE INFOCOM'06 conference held in a hotel in Barcelona in
2006~\cite{scott}.  A sample of $78$ participants from a range of
different companies and institutions were chosen and equipped with a
portable Bluetooth device, Intel iMote, able to detect similar devices
nearby.  Area ``inquiries'' were performed by the devices every 2
minutes, with a random delay or anticipation of 20 seconds.  The
delay/anticipation mechanism was implemented in order to avoid
synchronous measurements, because, while actively sweeping the area,
devices could not be detected by other devices.  A total number of
2730 distinct edges were recorded as active at least once in the
observation interval, while the number of edges active at a given time
is significantly lower, varying between 0 and 200, depending on the
time of the day.

\section{Activity-driven model}
\label{activity}

The activity-driven model, introduced in Ref.~\cite{Perra2012}, is a
simple model to generate time-varying graphs starting from the
empirical observation of the activity of each node, in terms of number
of contacts established per unit time. Given a characteristic
time-window $\Delta t$, one measures the activity potential $x_i$ of
each agent $i$, defined as the total number of interactions (edges)
established by $i$ in a time-window of length $\Delta t$ divided by
the total number of interactions established on average by all agents
in the same time interval. Then, each agent is assigned an activity
$a_i=\eta x_i$, which is the probability per unit time to create a new
connection or contact with any another agent $j$. The coefficient
$\eta$ is a rescaling factor, whose value is appropriately set in
order to ensure that the total number of active nodes per unit time in
the system is equal to $\eta\avg{x}N$, where $N$ is the total number
of agents. Notice that $\eta$ effectively determines the average
number of connections in a temporal snapshot whose length corresponds
to the resolution of the original data set.

The model works as follows. At each time $t$ the graph $G_t$ starts
with $N$ disconnected nodes. Then, each node $i$ becomes active with
probability $a_i\Delta t$ and connects to $m$ other randomly selected
nodes. At the following time-step, all the connections in $G_t$ are
deleted, and a new snapshot is sampled. 

Notice that time-varying graphs constructed through the
activity-driven model preserve the average degree of nodes in each
snapshot, but impose that connections have, on average, a duration
equal to $\Delta t$, effectively removing any temporal correlation
among edges. 

For the networks studied, we obtain mean raw activities $\avg{x}_{Infocom} \simeq  0.49$
and $\avg{x}_{Reality} \simeq 0.15$. 
Choosing a number $m=2$ of new links created for every activated node and constraining 
the average fraction of active nodes and the average number of contacts per node to be those of the real networks, 
 we obtain  $\eta_{Reality} \simeq 0.024$ and $\eta_{Infocom} \simeq 0.7$. 
Finally, the average activity of nodes becomes $\avg{a}_{Reality} = 0.004\pm 0.001$ and $\avg{a}_{Infocom} = 0.35 \pm 0.11$.

The aggregated versions of networks obtained from the activity-driven model were computed in two steps:
$i)$ a synthetic temporal network was created at the same temporal resolution  and of the same length as the 
original data set;  $ii)$ the synthetic network was aggregated on the appropriate time-window. 
This was done in order to mimic as closely as possible the procedure that we performed on the real networks, where a single temporal network was 
compared with its own aggregated versions.  

\begin{figure*}[!t]
  \begin{center}
    \includegraphics[width=6in]{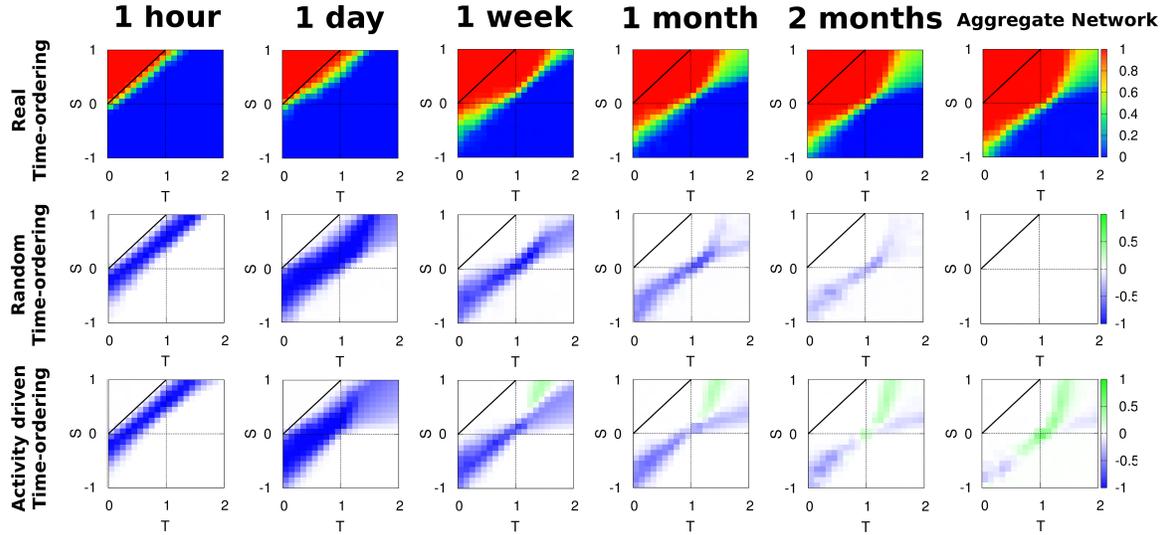}
  \end{center}
  \caption{(Color online). Here we report with a color-code the difference between the
    cooperation diagram corresponding to the original MIT Reality
    Mining graph sequence (top panels) and those obtained on the
    reshuffled (middle panels) and synthetic networks (bottom panels),
    at different values of the update interval $\Delta t$.}
  \label{fig:figs2}
\end{figure*}

\section{Temporal clustering}
\label{clustering}

Several metrics have been lately proposed to measure the tendency of
the edges of a time-varying graph to persist over time. One of the
most widely used is the unweighted temporal clustering, introduced in
Ref.~\cite{tang_sw}, which for a node $i$ of a time-varying graph is
defined as:
\begin{equation}
{\gamma}^i = \frac{1}{T-1} \sum_{t=1}^{T-1} \frac{\sum_j {a}^t_{ij}
  {a}_{ij}^{t+1}}{\sqrt{k_i^t k_i^{t+1}}}\,,
\end{equation}
where $a_{ij}^{t}$ are the elements of the adjacency matrix of the
time-varying graph at snapshot $t$, $k_i^{t}$ is the total number of
edges incident on node $i$ at snapshot $t$, and $T$ is the duration of
the whole observation interval. Notice that $\gamma^i$ takes values in
$[0,1]$. In general, a higher value of $\gamma^i$ is obtained when the
interactions of node $i$ persist longer in time, while $\gamma^i$
tends to zero if the interactions of $i$ are highly volatile.

If each snapshot of the time-varying graph is a weighted network,
where the weight $\omega^t_{ij}$ represents the strength if the
interaction between node $i$ and node $j$ at time $t$, we can define a
weighted version of the temporal clustering coefficient as follows:

\begin{equation}
{\gamma}_{w}^i = \frac{1}{T-1} \sum_{t=1}^{T-1} \frac{\sum_j
  {\omega}^t_{ij} {\omega}_{ij}^{t+1}}{s_i^t s_i^{t+1}}\,.
\end{equation}

Finally, if we focus more on the persistence of interaction strength
across subsequent network snapshots, we can define the extremal
temporal clustering as:
\begin{equation}
{\gamma}_{e}^i = \frac{1}{T-1} \sum_{t=1}^{T-1}  \frac{ \sum_j \min({\omega}^t_{ij}, {\omega}_{ij}^{t+1})}{\sqrt{s_i^t s_i^{t+1}}}\,,
\end{equation}
where by considering the minimum between ${\omega}^t_{ij}$ and
${\omega}_{ij}^{t+1}$ one can distinguish between persistent
interactions having constant strength over time and those interactions
having more volatile strength. As in our case social interactions are seen to be highly volatile in real data sets, the extremal version of the temporal clustering seems to be the best choice to unveil the persistence of social ties at short time scales.

\begin{figure*}[!t]
  \begin{center}
    \includegraphics[width=7in]{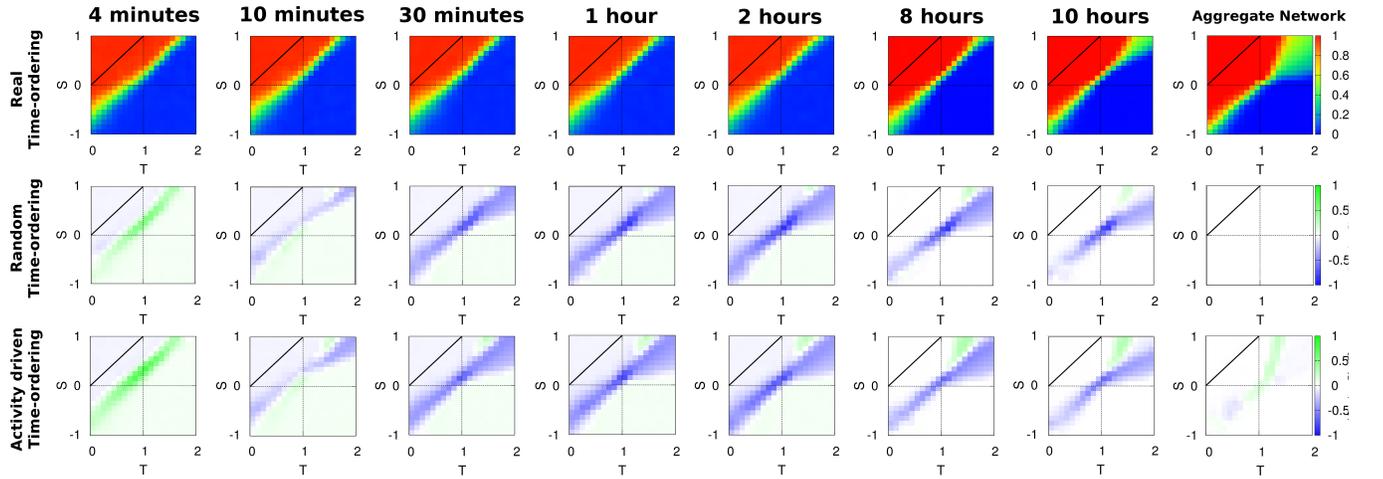}
  \end{center}
  \caption{(Color online). As in Fig.\ref{fig:figs2}, we report the differences between the
    cooperation diagram corresponding to the original INFOCOM'06 data set and those obtained on the
    reshuffled (middle panels) and synthetic networks (bottom panels),  at different values of the update interval $\Delta t$.}
  \label{fig:figs3}
\end{figure*}

\section{Differences in the Cooperation Diagrams}
\label{difference}

To highlight the effects that temporal correlations between pairwise interactions
have on the emergence of cooperation, we show in this appendix the quantitative differences among the cooperation level of the original data sets (as displayed by the fraction of cooperators as a function of  $T$ and $S$) and both their randomized and activity-driven versions. In Figs.~\ref{fig:figs2} and \ref{fig:figs3} we show these differences for the MIT Reality Mining and INFOCOM'06 graphs, respectively.

From these two figures, it becomes clear that most of the differences with the dynamics run on the original time-varying graph are concentrated at the interface between the regions in which 100\% of the nodes are cooperators [the red (light gray) areas in the top panels] and those where 100\% of the nodes become defectors [blue (dark gray) areas in the top panels]. Interestingly, for both data sets these differences are more pronounced for smaller values of $\Delta t$ and become less evident when $\Delta t$ increases, until they almost disappear for the largest aggregation interval.


\begin{thebibliography}{99}

\bibitem{penni05} E. Pennisi, ``How did cooperative behavior evolve?", {\em Science} {\bf 309}, 93 (2005).

\bibitem{penni09} E. Pennisi, ``On the origin of cooperation", {\em Science} {\bf 325}, 1196--1199 (2009).

\bibitem{maynard95} J. Maynard-Smith and E. Szathmary, ``The Major Transitions in Evolution" (Freeman, Oxford, UK, 1995).

\bibitem{nowak06} M.A. Nowak, ``Evolutionary Dynamics: Exploring the Equations of Life." (The Belknap Press of Harvard University Press, Cambridge, MA, 2006).

\bibitem{kollock98} P. Kollock, ``Social dilemmas: the anatomy of cooperation", {\em Annu. Rev. Sociol.} {\bf 24}, 183--214 (1998).

\bibitem{maynard73} J. Maynard-Smith, and G.R. Price, ``The logic of animal conflict", {\em Nature} {\bf 246} (5427), 15--18 (1973).

\bibitem{maynard82} J. Maynard-Smith, ``Evolution and the Theory of Games", Cambridge Univ. Press, Cambridge (UK), (1982).

\bibitem{gintis09} H. Gintis, ``Game Theory Evolving" (Princeton University Press, Princeton, NJ, 2009).

\bibitem{samuelson02} L. Samuelson, ``Evolution and game theory", {\em J. Econ. Perspect.} {\bf 16}, 47--66 (2002).






\bibitem{watts} D.J. Watts, and S.H. Strogatz, ``Collective dynamics of small-world networks" {\em Nature} {\bf 393}, 440--442 (1998).

\bibitem{strogatz} S.H. Strogatz, ``Exploring complex networks" {\em Nature} {\bf 410}, 268--276 (2001).
 
\bibitem{rev:albert} R. Albert, and A-L Barab\'asi, ``Statistical mechanics of complex networks",  {\em Rev. Mod. Phys.} {\bf 74}, 47--97 (2002).

\bibitem{rev:newman} M.E.J. Newman, ``The Structure and Function of Complex Networks" {\em SIAM Rev.} {\bf 45}, 167--256 (2003).

\bibitem{rev:bocc} S. Boccaletti, V. Latora, Y. Moreno, M. Chavez, and D-U Hwang, ``Complex Networks: Structure and Dynamics", {\em Phys. Rep.} {\bf 424}, 175--308 (2006).

\bibitem{rev:doro2} S.N. Dorogovtsev, A.V. Goltsev, and J.F.F. Mendes, ``Critical phenomena in complex networks", {\em Rev Mod. Phys.} {\bf 80}, 1275--1335 (2008).

\bibitem{rev:castellano} C. Castellano, S. Fortunato, and V. Loreto, ``Statistical physics of social dynamics", {\em Rev. Mod. Phys.} {\bf 81} 591--646 (2009).


\bibitem{pachprl} F.C. Santos, and J.M. Pacheco, ``Scale-Free Networks Provide a Unifying Framework for the Emergence of Cooperation", {\em Phys. Rev. Lett.} {\bf 95}, 098104 (2005).

\bibitem{pachpnas} F.C. Santos, J.M. Pacheco, and T. Lenaerts, ``Evolutionary dynamics of social dilemmas in structured heterogeneous populations", {\em Proc. Natl. Acad. Sci. (U.S.A.)} {\bf 103}, 3490--3494 (2006).

\bibitem{prl} J. G\'omez-Garde\~{n}es, M. Campillo, L.M. Flor\'{\i}a, and Y. Moreno, ``Dynamical Organization of Cooperation in Complex Topologies",  {\em Phys. Rev. Lett.} {\bf 98}, 108103 (2007).

\bibitem{assenza} S. Assenza, J. G\'omez-Garde\~nes, and V. Latora, ``Enhancement of cooperation in highly clustered scale-free networks", {\em Phys. Rev. E} {\bf 78}, 017101 (2008).

\bibitem{pachnature} F.C. Santos, M.D. Santos, and J.M. Pacheco, ``Social diversity promotes the emergence of cooperation in public goods games", {\em Nature} {\bf 454}, 213--216 (2008).
 
\bibitem{chaosPGG} J. G\'omez-Garde{\~n}es, M. Romance, R. Criado, D. Vilone, and A. S\'anchez, ``Evolutionary games defined at the network  mesoscale: The Public Goods game", {\em Chaos} {\bf 21}, 016113 (2011).



\bibitem{percpggrev} M. Perc, J. G\'omez-Garde\~nes, A. Szolnoki, L. M. Floria, and Y. Moreno, ``Evolutionary dynamics of group interactions on structured populations: a review", {\em J. Roy. Soc. Interface} {\bf 10}, 20120997 (2013).

\bibitem{szaborev} G. Szab\'o, and G. F\'ath, ``Evolutionary games on graphs", {\em Phys. Rep.} {\bf 446}, 97 (2007).

\bibitem{jackson08} M.D. Jackson, ``Social and Economic Networks" (Princeton Univ. Press, Princeton, NJ, 2008).

\bibitem{anxorev} C.P. Roca, J. Cuesta, and A. S\'anchez, ``Evolutionary game theory: temporal and spatial effects beyond replicator dynamics", {\em Phys. Life Rev.} {\bf 6}, 208 (2009).

\bibitem{percrev} M. Perc, and A. Szolnoki, ``Coevolutionary games - A mini review", {\em BioSystems} {\bf 99}, 109--125 (2010).

\bibitem{grossrev} T. Gross, and B. Blasius, ``Adaptive coevolutionary networks: a review", {\em J. R. Soc. Interface} {\bf 5}, 259--271 (2008).



\bibitem{eaglepentland} N. Eagle and A. Pentland, ``Reality mining: sensing complex social systems", {\em Person. Ubiq. Comput.} {\bf 10}, 255--268 (2006).

\bibitem{scott} J. Scott \etal, ``CRAWDAD Trace", INFOCOM, Barcelona (2006).

\bibitem{isella} L. Isella, M. Romano, A. Barrat, C. Cattuto, V. Colizza, \etal, ``Close Encounters in a Pediatric Ward: Measuring Face-to-Face Proximity and Mixing Patterns with Wearable Sensors." {\em PLoS ONE} {\bf 6}, e17144 (2011).

\bibitem{sthele_primary} J. Stehle, N. Voirin, A. Barrat, C. Cattuto, L. Isella, \etal ``High-Resolution Measurements of Face-to-Face Contact Patterns in a Primary School." {\em PLoS ONE} {\bf 6}, e23176 (2011). 


\bibitem{cattuto} L. Isella, J. Stehl\'e, A. Barrat, C. Cattuto, J-F Pinton, and W. Van den Broeck, ``What's in a crowd? Analysis of face-to-face behavioral networks", {\em J. Theor. Biol.} {\bf 271}, 166--180 (2011).

\bibitem{saramaki} M. Karsai, M Kivel\"a, R.K. Pan, K. Kaski, J. Kert\'esz, A-L Barab\'asi, and J. Saram\"aki, ``Small But Slow World: How Network Topology and Burstiness Slow Down Spreading", {\em Phys. Rev. E} {\bf 83}, 025102(R) (2011).



\bibitem{barabasi2005}
A-L Barab\'asi, ``The origin of bursts and heavy tails in human dynamics" {\em Nature} {\bf 435}, 207--211 (2005).

\bibitem{bianconi2} J. Stehl\'e, A. Barrat, and G. Bianconi, ``Dynamical and bursty interactions in social networks", {\em Phys. Rev. E} {\bf 81}, 035101(R) (2010).



\bibitem{gonzales} M.C. Gonz\'alez, C.A. Hidalgo, and A-L Barab\'asi, ``Understanding individual human mobility patterns" {\em Nature} {\bf 453}, 779--782 (2008).

\bibitem{szell} M. Szell, R. Lambiotte, and S. Thurner, ``Multi-relational Organization of Large-scale Social Networks in an
  Online World",  {\em Proc. Natl. Acad. Sci. (U.S.A.)} {\bf 107}, 13636--13641 (2010).

\bibitem{sinatra} M. Szell, R. Sinatra, G. Petri, S. Thurner, and V. Latora, ``Understanding mobility in a social petri dish", {\em Sci. Rep.} {\bf 2}, 457 (2012).




\bibitem{kossinets} G. Kossinets, J. Kleinberg, and D. Watts, ``The Structure of Information Pathways in a Social Communication Network,"
  Proceedings of the 14th ACM SIGKDD International Conference on Knowledge Discovery and Data Mining (ACM, New York, 2008).

\bibitem{kostakos} V. Kostakos, ``Temporal graphs", {\em Physica A} {\bf 388}, 1007--1023 (2009).

\bibitem{tang_distance} J. Tang, M. Musolesi, C. Mascolo, and
  V. Latora, ``Temporal Distance Metrics for Social Network Analysis",
  Proceedings of the 2nd ACM SIGCOMM Workshop on Online Social
  Networks (WOSN'09) (ACM, New York, 2009).

\bibitem{holme_review} P. Holme and J. Saram\"aki, ``Temporal networks", {\em Phys. Rep.} {\bf 519}, 97--125 (2012).



\bibitem{tang_sw} J. Tang, S. Scellato, M. Musolesi, C. Mascolo, and
  V. Latora, ``Small-world behavior in time-varying graphs."
  {\em Phys. Rev. E} {\bf 81}, 055101(R) (2010).

\bibitem{nicosia_chapter} V. Nicosia, J. Tang, C. Mascolo,
  M. Musolesi, G. Russo, V. Latora "Graph Metrics for Temporal
  Networks", edited by P. Holme and J. Saram\"aki, Temporal
  Networks. (Springer, Berlin, 2013), pp. 15-40.

\bibitem{pan_paths} R.K. Pan, and J. Saram\"aki, ``Path lengths, correlations, and centrality in temporal networks," {\em Phys. Rev. E} {\bf 84}, 016105 (2011).

\bibitem{kovanen_motifs} L. Kovanen, M. Karsai, K. Kaski, J. Kertesz, and J. Saram\"aki. ``Temporal motifs in time-dependent networks" {\em J. Stat. Mech.}, P11005 (2011).

\bibitem{tang_centrality} J. Tang, M. Musolesi, C. Mascolo, V. Latora,
  and V. Nicosia, ``Analysing Information Flows and Key Mediators
  through Temporal Centrality Metrics" Proceedings of the 3rd ACM
  Workshop on Social Network Systems (SNS'10), (ACM, New York, 2010).

\bibitem{nicosia_components} V. Nicosia, J. Tang, M. Musolesi,
  G. Russo, C. Mascolo, and V. Latora, ``Components in time-varying
  graphs", {\em Chaos} {\bf 22}, 023101 (2012).

\bibitem{mucha10} P.J. Mucha, T. Richardson, K. Macon, M.A. Porter,
  and J-P Onnela, ``Community structure in time-dependent, multiscale,
  and multiplex networks", {\em Science} {\bf 328}, 876--878 (2010).



\bibitem{starnini_rw} M. Starnini, A. Baronchelli, A. Barrat, and
  R. Pastor-Satorras, ``Random walks on temporal networks",
  {\em Phys. Rev. E} {\bf 85}, 056115 (2012).

\bibitem{perra_walking} N. Perra, A. Baronchelli, D. Mocanu,
  B. Gon\c{c}alves, R. Pastor-Satorras, and A. Vespignani, ``Walking
  and searching in time-varying networks", {\em Phys. Rev. Lett.} {\bf 109},
  238701 (2012).

\bibitem{ribeiro2013} B. Ribeiro, N. Perra, and A. Baronchelli, ``Quantifying the effect of temporal resolution on time-varying networks." {\em Sci. Rep.} {\bf 3}, 1 (2013).

\bibitem{rocha_1} L.E.C. Rocha, F. Liljeros, and P. Holme, ``Simulated
  Epidemics in an Empirical Spatiotemporal Network of 50,185 Sexual
  Contacts."  {\em PLoS Comp. Biol.} {\bf 7}, e1001109 (2011).

\bibitem{rocha_2} L.E.C. Rocha, A. Decuyper, and V.D. Blondel, ``Epidemics on a stochastic model of temporal network",  {\em Dynamics on and of Complex Networks, Volume 2. Modeling and Simulation in Science, Engineering and Technology} (Springer, New York, 2013), pp. 301--314.

\bibitem{rocha_3} L.E.C. Rocha and V.D. Blondel, ``Bursts of vertex activation and epidemics in evolving networks." {\em PLoS Comp. Biol.} {\bf 9}, e1002974 (2013).

\bibitem{diazguilera} N. Fujiwara, J. Kurths, and A. D\'iaz-Guilera, ``Synchronization in networks of mobile oscillators." {\em Phys. Rev. E} {\bf 83}, 025101(R) (2011).


\bibitem{fermi1} L.E. Blume, ``The Statistical Mechanics of Strategic Interaction." {\em Games Econ. Behav.} {\bf 5}, 387 (1993).

\bibitem{fermi2} G. Szab\'o and C. T\"oke, ``Evolutionary prisoner's dilemma game on a square lattice", {\em Phys. Rev. E} {\bf 58}, 69 (1998).
  
\bibitem{Perra2012} N. Perra, B. Gon\c{c}alves, R. Pastor-Satorras, and A. Vespignani, ``Activity driven modeling of time varying networks'', {\em Sci. Rep.} {\bf 2}, 469 (2012).


\bibitem{wang_scirep_2012} Z. Wang, A. Szolnoki, and M. Perc, ``If players are sparse social dilemmas are too: Importance of percolation for evolution of cooperation.''  {\em Sci. Rep.} {\bf 2}, 369 (2012).

\bibitem{cuesta} C.P. Roca, J.A. Cuesta, and A. S\'anchez, ``Time Scales in Evolutionary Dynamics'', {\em Phys. Rev. Lett.} {\bf 97}, 158701 (2006).

\bibitem{pachtraul} J.M. Pacheco, A. Traulsen, and M.A. Nowak, ``Coevolution of Strategy and Structure in Complex Networks with Dynamical Linking'', {\em Phys. Rev. Lett.} {\bf 97}, 258103 (2006).



  
  
\end{thebibliography}
\end{document}